\documentstyle[11pt,aaspp4]{article}
\def	\cm	{{\rm\,cm}}
\def	\erg	{{\rm\,ergs}}

\def	\eV	{{\rm\,eV}}
\def	\K	{{\rm\,K}}
\def	\Qabs	{Q_{\rm UV}}	
\def	\Rion	{R_{\rm ion}}
\def	\s	{{\rm\,s}}
\def\plotBTD#1#2{%
  \expandafter\ifx\csname epsfbox\endcsname\relax
    \immediate\write16{%
        You need to input epsf; I'll do it for you
    }%
    \input epsf
  \fi
  \epsfysize=#2
     \openin 1 #1 \ifeof 1
        \immediate\write16{Can't open #1}%
        \vskip \the\epsfysize
      \else
         \closein 1
         \centerline{\epsfbox{#1}}%
      \fi
}

\begin{document}
\title{Dust sublimation by GRBs and its implications}
\author{E. Waxman$^1$ and B. T. Draine$^2$}
\affil{$^1$Department of Condensed-Matter Physics, Weizmann Institute,
Rehovot 76100, Israel}
\affil{$^2$Princeton University Observatory, Peyton Hall, Princeton, NJ 08544}

\begin{abstract}

The prompt optical flash recently detected accompanying GRB990123
suggests that, for at least some GRBs,
$\gamma$-ray emission 
is
accompanied by prompt optical-UV emission with luminosity
$L(1-7.5\eV)
\approx1\times10^{49}(\Delta\Omega/4\pi){\rm erg/s}$,
where $\Delta\Omega$ is the solid angle into which $\gamma$-ray and 
optical-UV emission is beamed. 
Such an optical-UV flash can destroy dust {\it in the beam} by sublimation
out to an appreciable distance, 
$\approx10$~pc,
and may clear the dust out of 
as much as
$\sim10^7(\Delta\Omega/4\pi) M_\odot$ 
of molecular cloud material
on 
an apparent
time scale of $\sim$ ten seconds. Detection of time dependent extinction
on this time scale would therefore provide strong constraints on the GRB source
environment. Dust destruction implies that existing, or
future, observations of not-heavily-reddened fireballs are not
inconsistent with GRBs being associated with star forming regions.
In this case, however, if $\gamma$-ray emission is 
highly beamed, the expanding fireball would become reddened on 
a $\sim$1 week time scale.

If the optical depth due to dust 
beyond $\approx8$~pc from the GRB
is $0.2\lesssim\tau_V\lesssim2$, most of the UV flash energy is 
converted to infra-red, $\lambda\approx1\mu$m, radiation with luminosity 
$L_{\rm IR}\approx10^{41}{\rm erg/s}$ extending over
an apparent duration of $\approx20(1+z)(\Delta\Omega/0.01)$~day. 
Dust infra-red emission may 
already have been observed in GRB970228 and GRB980326, and may possibly 
explain their unusual late time behavior.

\end{abstract}

\section{Introduction}

A prompt, 9-th magnitude, optical flash has recently been detected 
(\cite{Akerlof99}) accompanying 
the gamma ray burst (GRB) GRB990123.
The most natural explanation of this flash
is emission from a reverse shock propagating into fireball ejecta shortly after
it interacts with surrounding gas (\cite{SP99b,MR99}). Although optical-UV
emission from the reverse shock accompanying, or following shortly after, the 
$\gamma$-ray emission has been predicted (\cite{MRP94,MRAG,PM98,SP99a})
based on the simplest fireball models for gamma-ray bursts (GRBs), 
the intensity
of optical emission could not have been reliably predicted due to the
uncertainty in reverse shock parameters.
The observations of GRB990123
suggest that the electron and magnetic field energy fractions
in the reverse shock are similar to those in
the forward shock propagating
into the surrounding gas and producing the long term afterglow. This in turn
implies that strong optical flashes accompanying $\gamma$-ray emission is
a generic GRB characteristic. This is consistent with previous non-detection
of optical flashes (\cite{LOTIS}) 
given the wide GRB luminosity function (\cite{KTH98,MnM98,HnF98}) 
and the fact that 
GRB990123 is in the top 0.3\% of the BATSE brightness distribution
(\cite{Kouv99}). 

There is evidence in several GRB afterglows for significant dust extinction,
which may imply an association of GRBs with star forming regions
(\cite{SFR1,SFR2,SFR3}). 
It has been shown that if GRBs indeed reside in such environment,
the ionizing X-ray and UV afterglow radiation may lead to time dependent
(on hour time scale)
absorption (\cite{Perna98}) and emission (\cite{Ghisellini98,Bottcher98}) 
line features. 
If H$_2$ is present near the GRB, conspicuous 1110--1650\AA absorption 
will be imprinted on the spectrum, followed by UV fluorescence
(\cite{D99b}).
On longer time scales, up to $\sim10^5$~yr, GRB photo-ionization
may lead to indicative recombination line features, which may
allow identification of GRB remnants in nearby galaxies 
(\cite{Loeb98,Perna99}).
Here, we discuss dust sublimation by the optical-UV 
flash accompanying the GRB.
Since GRB990123 was an exceptionally bright burst, with
exceptionally high intrinsic $\gamma$-ray luminosity,
we first discuss in \S2 the optical-UV flash itself to obtain the scaling
of optical-UV luminosity with burst energy,
and to estimate the prompt optical-UV emission at energies
above the $2(1+z)$~eV energy of the prompt
emission observed by Akerlof et al. 
We show that if electron and 
magnetic field energy fractions in the reverse shock are similar among 
different bursts, then an optical-UV flash of luminosity 
$L_{\rm UV}\approx3\times10^{49}(\Delta\Omega/4\pi){\rm erg/s}$,
where $\Delta\Omega$ is the beaming solid angle,
is expected for a typical GRB.
Dust destruction 
by thermal sublimation is discussed in \S3, 
and other possible destruction
mechanism are considered in \S4.
It should be emphasized that 
the analysis of 
the physics of
dust destruction is independent of the model for optical-UV emission, and
therefore of the discussion presented in \S2.
However, we present numerical results for dust destruction based on
the characteristic optical-UV luminosity derived in \S2.
In \S5 we discuss dust infra-red emission.
The implications of our results are discussed in \S6.

\section{The Prompt Optical-UV Luminosity}
\subsection{Fireball Dynamics}

In fireball models of GRBs (see \cite{Piran98} for a recent review), 
the energy released by an explosion
is converted to kinetic energy of a thin baryonic shell expanding at an
ultra-relativistic speed. After producing the GRB, the shell impacts on
surrounding gas, driving an ultra-relativistic shock into the ambient
medium. After a short transition phase, the expanding blast wave approaches
a self-similar behavior (\cite{BnM}),
where the expansion Lorentz factor decreases with
radius as $\Gamma\propto r^{-3/2}$.   
The initial interaction of fireball ejecta with surrounding gas 
produces a 
reverse shock which propagates into 
and decelerates the fireball ejecta.
Transition to self-similar behavior occurs on a time scale comparable
to the reverse shock crossing time of the ejecta. 

The long term afterglow is produced by the forward, expanding shock that 
propagates into the surrounding gas. This shock continuously heats
fresh gas and accelerates relativistic electrons, which produce the
observed radiation through synchrotron emission.
The most natural explanation of the optical flash is that it is 
due to synchrotron emission of electrons which are accelerated by the reverse 
shock, during the transition to self-similar behavior. Once the reverse
shock crosses the ejecta, the ejecta 
expand and cool
adiabatically. Thus,
emission from the fireball ejecta is suppressed after the transition
to self-similar expansion.

Since the optical flash is produced when the reverse shock crosses the ejecta,
the plasma emitting the radiation expands with a Lorentz factor which
is close to that given by the Blandford-McKee self-similar solution,
$\Gamma=(17E/16\pi n m_p c^2)^{1/2}r^{-3/2}$, where $E$ is the fireball energy
and $n$ is the surrounding gas number density. 
The characteristic time 
over which radiation emitted by the fireball at radius $r$ 
is observed by a distant observer is $\Delta t\approx r/4\Gamma^2 c$ 
(\cite{Wring}). 
The 
plasma Lorentz factor 
during optical flash emission
\begin{equation}
\Gamma\simeq\left({17E\over 1024\pi n m_p c^5}\right)^{1/8}
\left({\Delta t^{\rm ob.}\over 1+z}\right)^{-3/8}=345\left({E_{53}\over 
n_0}\right)^{1/8}\left({2.5\over1+z}\Delta t^{\rm ob.}_{1}\right)^{-3/8},
\label{Gamma}
\end{equation}
where $E=10^{53}E_{53}$~erg, $\Delta t^{\rm ob.}=10
\Delta t^{\rm ob.}_{1}$~s 
is the observed duration, $\Delta t^{\rm ob.}=(1+z)\Delta t$,
and $n=1n_0\cm^{-3}$. 

Transition to self-similar expansion occurs on time scale $\Delta t^{\rm ob.}$
comparable to the longer of the two time scales set by the initial
conditions: the (observed) GRB duration $\Delta t_{\rm GRB}$ and the 
(observed) time
$\Delta t_{\Gamma}$ at which the self-similar Lorentz factor
equals the original ejecta Lorentz factor $\Gamma_i$, 
$\Gamma(\Delta t^{\rm ob.}=\Delta t_\Gamma)=\Gamma_i$. That is,
\begin{equation}
\Delta t^{\rm ob.}=\max\left[\Delta t_{\rm GRB}, 15{1+z\over2.5}
\left({E_{53}\over
n_0}\right)^{1/3}\left({\Gamma_i\over300}\right)^{-8/3}{\rm\, s}\right].
\label{Delta_t}
\end{equation}
During the transition, the unshocked fireball
ejecta propagate at the original expansion Lorentz factor, $\Gamma_i>\Gamma$,
and the Lorentz factor of plasma shocked by the reverse shock in
the rest frame of the unshocked ejecta is $\simeq\Gamma_i/\Gamma$.
If $\Delta t^{\rm ob.}\simeq\Delta t_{\rm GRB}\gg\Delta t_{\Gamma}$ then 
$\Gamma_i/\Gamma\gg1$, the reverse shock is relativistic,
and the Lorentz factor associated with the random motion of protons 
in the reverse shock is $\gamma_p^R\simeq\Gamma_i/\Gamma$. 
If $\Delta t^{\rm ob.}\simeq\Delta t_{\Gamma}\gg\Delta t_{\rm GRB}$ then 
$\Gamma_i/\Gamma\sim1$, and the reverse shock is not relativistic. 
Nevertheless, the following argument suggests that the reverse shock
speed is not far below $c$, and that the protons are therefore 
heated to 
relativistic energy, $\gamma_p^R-1\simeq1$. The comoving time, measured
in the fireball ejecta frame prior to deceleration, 
is $t_{\rm co.}\simeq r/\Gamma_i c$. The expansion Lorentz
factor is expected to vary across the ejecta, $\Delta \Gamma_i/\Gamma_i\sim1$.
Such variation would lead
to expansion of the ejecta, in the comoving frame, at relativistic speed. 
Thus, at the deceleration radius, $t_{\rm co.}\simeq\Gamma_i\Delta t$,
the ejecta width exceeds $\simeq ct_{\rm co.}\simeq\Gamma_i c \Delta t$. Since
the reverse shock should cross the ejecta over a deceleration time 
scale, $\simeq\Gamma \Delta t$, the reverse shock speed must be close to 
$c$. We therefore conclude that 
the Lorentz factor associated with the random motion of protons 
in the reverse shock is approximately given by 
$\gamma_p^R-1\simeq \Gamma_i/\Gamma$ for both $\Gamma_i/\Gamma\sim1$
and $\Gamma_i/\Gamma\gg1$. For protons shocked by 
the forward shock $\gamma_p^F\simeq\Gamma$, and therefore the ratio
between thermal energy per proton in the reverse and forward shocks
is $(\gamma_p^R-1)/(\gamma_p^F-1)\simeq\Gamma_i/\Gamma^2$.
Below we use this relation to
derive the emission characteristics of the reverse shock by scaling
the exact analytic results given for the forward shock emission 
in \cite{GnW99}.

\subsection{Optical-UV Emission}

If the fraction of thermal
energy carried by electrons, $\xi_e$, and magnetic field, $\xi_B$, is similar
in the forward and reverse shocks, then the frequency of peak synchrotron
emission from the reverse shock is smaller than that of the
forward shock by a factor $\simeq(\Gamma_i/\Gamma^2)^2$.  
This is due to the fact that the energy density behind the reverse and forward
shocks are similar, so that similar $\xi_B$ implies similar magnetic
field strength in both regions, while similar $\xi_e$ implies 
$\gamma_e^R/\gamma_e^F\simeq
(\gamma_p^R-1)/(\gamma_p^F-1)\simeq\Gamma_i/\Gamma^2$.
Using Eq. (10) of \cite{GnW99} for the forward shock peak 
frequency, we find
that the reverse shock 
emission peaks at a frequency 
\begin{equation}
\nu_m \approx 1.7\times10^{14}
\left({\xi_e\over0.2}\right)^2
      \left({\xi_B\over0.01}\right)^{1/2}\left({\Gamma_i\over300}\right)^2
  n_0^{1/2}{\rm Hz}
\label{nu_m}
\end{equation}
(measured at the GRB redshift).
Here $\xi_e$ and $\xi_B$ are the values relevant for the reverse shock,
which in general 
may differ from those of the forward shock. The numerical values
we have used are those characteristic of the forward shock 
(\cite{W97a,W97b,WG98,GPS}). 

As demonstrated below, the cooling time of electrons in the reverse
shock, radiating in the optical-UV range, is long compared to the fireball 
expansion time. In this case,
the peak synchrotron intensity is proportional to the product of
magnetic field strength and number of radiating electrons (and independent
of the electron Lorentz factor). 
The number of radiating electrons in the
reverse shock is larger than 
in the forward shock
by a factor $\Gamma^2/\Gamma_i$. This can be 
deduced from the following considerations. The proton random Lorentz
factor in the reverse shock is $\Gamma^2/\Gamma_i$ times smaller than that
in the forward shock. Since the energy density in both regions is similar,
the density of protons, and therefore of electrons, in the reverse shock is 
higher than that in the forward shock by the same factor. In addition,
the width of the shocked fireball ejecta is similar to that of the shell
of shocked surrounding gas, since both shocks propagate relativistically
in the shocked plasma frame. Thus, if $\xi_B$ is similar in the
reverse and forward shock, the peak synchrotron intensity 
$f_m$
is higher in the
reverse shock by a factor $\Gamma^2/\Gamma_i$. Using Eq. (11) of
\cite{GnW99} we find for the reverse shock
\begin{equation}
f_m \approx 1 h_{65}^2
    \left({\sqrt{2.5}-1\over\sqrt{1+z}-1}\right)^2
    \left({\xi_B\over0.01}\right)^{1/2}\left({\Gamma_i\over300}\right)^{-1}
n_0^{1/4}
E_{53}^{5/4}
\left({2.5\over1+z}\Delta t^{\rm ob.}_{1}\right)^{-3/4}
{\rm\ Jy}
\label{f_m}
\end{equation}
for a
flat universe with zero cosmological constant, and $H_0=65h_{65}
{\rm km/s\ Mpc}$. 

The decay of GRB990123 optical flux implies an electron
energy distribution $dN_e/dE_e\propto E_e^{-p}$ with $p\approx2$ 
(\cite{MR99}), for which the intensity at $\nu>\nu_m$ is 
$f_\nu\propto\nu^{-1/2}$. Thus, the 
optical [$\nu=(5(1+z)\times10^{14}$~Hz]
intensity is
\begin{equation}
f_V \approx 0.4 h_{65}^2\left({2.5\over1+z}\right)^{1/2}
    \left({\sqrt{2.5}-1\over\sqrt{1+z}-1}\right)^2\left({\xi_e\over0.2}\right)
    \left({\xi_B\over0.01}\right)^{3/4}
n_0^{1/2}
E_{53}^{5/4}
\left({2.5\over1+z}\Delta t^{\rm ob.}_{1}\right)^{-3/4}
{\rm\ Jy}.
\label{eq:f_V}
\end{equation}
For the parameters of GRB990123, 
$\Delta t^{\rm ob.}
\approx60$~s and $E\approx10^{54}$~erg 
(based on the $\gamma$-ray fluence $\sim3\times10^{-4}{\rm erg}\cm^{-2}$),
we obtain $f_V\approx2$~Jy, approximately twice the 
peak
observed flux.
Thus, the observed
optical flash can be naturally explained by the simplest fireball model,
provided the reverse shock parameters $\xi_e$ and $\xi_B$ are similar
to those implied for the forward shock by afterglow observations,
$\xi_e\approx0.2$ and $\xi_B\approx0.01$. If this is 
typical,
than for a typical GRB, with $E\sim10^{53}$~erg and 
$\Delta t^{\rm ob.}\approx10$~s,
we find
\begin{equation}
L_\nu={4\pi d_L^2\over 1+z} f_\nu \approx 10^{34}
    \left({\xi_e\over0.2}\right)\left({\xi_B\over0.01}\right)^{3/4}
n_0^{1/2}
E_{53}^{5/4}\left({2.5\over1+z}
\Delta t^{\rm ob.}_{1}
\right)^{-3/4}
    \left({\nu\over10^{15}{\rm Hz}}\right)^{-1/2}{\rm\ erg/s\ Hz}.
\label{L_nu}
\end{equation}
where $\nu$ is the frequency at the redshift of the GRB.
The form (\ref{L_nu}) is valid for frequencies 
$\nu$
larger than the peak
frequency $\nu_m$, and smaller than the frequency $\nu_c$
above which
emission is
dominated by electrons for which the cooling time is shorter than the
dynamic time. Since the energy density in the reverse and forward shock
regions is similar, $\nu_c$ (at the GRB redshift)
is given by
\begin{equation}
\nu_c\sim 10^{17}\left({\xi_B\over0.01}\right)^{-3/2}
n_0^{-1}
E_{53}^{-1/2}\left({2.5\over1+z}
\Delta t_{1}^{\rm ob.}
\right)^{-1/2}
  {\rm Hz}.
\label{nu_c}
\end{equation}
At frequencies $\nu>\nu_c$ the spectrum steepens to $L_\nu\propto\nu^{-1}$.

Strong optical-UV emission requires, similar to 
GRB $\gamma$-ray production, large initial Lorentz factor, 
$\Gamma_i>100$: 
Eq. (\ref{Delta_t}) implies that $\Delta t\propto\Gamma_i^{-8/3}$ 
for $\Gamma_i\lesssim300$, 
and the optical-UV luminosity given by 
Eq. (\ref{L_nu}) drops as $L_\nu\propto\Gamma_i^{2}$. This implies that the 
plasma emitting the optical-UV photons must also be expanding at high
Lorentz factor, $\Gamma>100$ [see Eqs. (\ref{Delta_t},\ref{Gamma})], 
and strong optical-UV emission may be confined,
similar to $\gamma$-rays, to a cone
around the line of sight of opening angle 
$1\gg\theta_\gamma>1/\Gamma\sim0.01$.
This may be the case if, e.g.,
the fireball is a jet of opening angle $\theta_j=\theta_\gamma$.
A jet of finite opening angle expands as if it were a conical section of
a spherical fireball, as long as $\theta_j>1/\Gamma$. Thus, the analysis
presented above is valid for a jet-like fireball. In this case, the
energy $E$ in the above equations should be understood as the energy
that the fireball would have carried if it were spherically symmetric,
and the optical-UV, as well as $\gamma$-ray, emission is confined 
to a small solid angle $\Delta\Omega=\pi\theta_j^2$ with optical-UV luminosity 
given by $\Delta\Omega L_\nu$. 

It should be noted that prompt optical emission has been observed for only
1 GRB to date, and therefore may not be typical of GRBs.
Indeed, fireballs in low density environments with $n\ll 1\cm^{-3}$ would
not be expected to produce strong prompt emission [see Eq. (\ref{eq:f_V})].
However, we note from Eq. (\ref{eq:f_V}) that optical flashes are expected
for typical GRBs with $E_{53}\approx 1$ and 
$n_0\gtrsim 1$, 
and should not be limited to
unusually luminous GRBs such as GRB990123.

\section{Dust Destruction by Thermal Sublimation}

As discussed in \S3.3 below, in dense regions
radiation with $h\nu>13.6$~eV will largely go into photoionizing
H and H$_2$, and photons in the 7.5--13.6~eV range will primarily
be absorbed by H$_2$, which is rovibrationally
excited by ultraviolet pumping (Draine 1999b).
We therefore first discuss, in \S3.1 and \S3.2, sublimation of 
dust grains under the assumption that dust heating is
dominated by $1{\rm\,eV}<h\nu<7.5$~eV photons, and
discuss the contribution of 7.5--50~eV photons to dust sublimation
in \S3.3.
Using equations (\ref{L_nu}) and (\ref{nu_c}), 
we find that for typical GRB parameters,
the prompt luminosity in the 1--7.5~eV range is
$2\times10^{49}~{\rm erg/s}$ and the 7.5--50~eV prompt
luminosity is $\sim 5\times 10^{49}~{\rm erg/s}$.
Since our simple analysis overestimates the flux of GRB990123 by a factor
$\sim2$, we will take the typical 1--7.5~eV luminosity to be
$L_{1-7.5}=10^{49}L_{49}{\rm erg/s}$, 
and the 7.5--50~eV luminosity to be
$L_{7.5-50}=2.5\times10^{49}L_{49}~{\rm erg/s}$.

\subsection{Grain Heating}

Consider a grain of radius $a$ located at a distance $r$ from a transient
source of radiation radiating a 
1--7.5 eV power $L_{1-7.5}(\Delta\Omega/4\pi)$
into a solid angle $\Delta\Omega$.
If the radiation from the GRB is ``beamed'' into $\Delta\Omega<4\pi$,
we will consider only dust grains within the beam.
Note that we expect the beaming of optical-UV emission from the forward
shock to be similar to that of gamma-ray emission.

The grain temperature $T$ is determined by
\begin{equation}
e^{-\tau}\frac
{L_{1-7.5}}
{4\pi r^2} 
\Qabs
\pi a^2
=
\langle 
Q
\rangle_T 4\pi a^2 \sigma T^4 -
4\pi a^2 \frac{da}{dt} \frac{\rho}{m} B \quad,
\label{eq:balance}
\end{equation}
where 
$\tau$ is the effective optical depth for attenuation of the optical-UV flash,
$\rho$ is the density of the grain material, $m$ is the
mean atomic mass, 
$B$ is the chemical binding energy per atom,
\begin{equation}
\langle 
Q
\rangle_T 
\equiv
\frac{\int B_\nu(T) Q_{{\rm abs},\nu} d\nu}
{\int B_\nu(T) d\nu} ~~
\end{equation}
is the usual Planck-averaged absorption efficiency,
and
$\Qabs$ is the absorption efficiency factor averaged over the
1--7.5~eV spectrum of the optical-UV flash.
For the grain radii $a\gtrsim 10^{-5}\cm$ expected in dense clouds,
we will assume $\Qabs\approx 1.$
Since we are interested in energy depositions large enough to sublime
grains, the heat capacity of the grain has been neglected 
in equation (\ref{eq:balance}).

The sublimation rate can be approximated by
\begin{equation}
\frac{da}{dt} = - \left(\frac{m}{\rho}\right)^{1/3} \nu_0 e^{-B/kT} ~.
\end{equation}
Guhathakurta \& Draine (1989) have estimated
$\nu_0\approx 2\times10^{15}{\rm s}^{-1}$, 
$B/k=68100{\rm K}$ for Mg$_2$SiO$_4$, and $\nu_0=2\times10^{14}{\rm s}^{-1}$,
$B/k=81200{\rm K}$ for graphite.
We adopt $\nu_0=1\times10^{15}{\rm s}^{-1}$, $B/k=7\times10^4{\rm K}$, 
and $\rho/m\approx 10^{23}\cm^{-3}$
as representative values for refractory grains.
If we assume the grain temperature $T$ is approximately constant over
the time 
$\Delta t = \Delta t^{\rm ob.}/(1+z)$,
then the condition for the grain to be completely
sublimed during this time would be $T > T_c$, where
\begin{equation}
T_c = \frac{B/k}{
\ln\left[\left(m/\rho\right)^{1/3}\left(\nu_0\Delta t/a\right)\right]
	}
\approx 2300 {\rm K} 
\left[ 1+
0.033\ln\left(
	\frac{a_{-5}}
{\Delta t_{1}}
	\right)
\right]
\label{eq:T_c}
\end{equation}
where $a_{-5}\equiv a/10^{-5}\cm$.  Equivalently, the grain survival
time at temperature $T$ is
\begin{equation}
t_{\rm surv}(T) = {a\over |da/dt|} = 7.7 a_{-5} 
\exp\left[7\times10^4\K\left({1\over T}-{1\over 2300\K}\right)\right]{\rm\, s}
\quad .
\label{eq:t_sur}
\end{equation}

The infrared emissivity is quite different for graphite and silicate
materials (Draine \& Lee 1984; Draine 1999a). For the temperature range
of interest for dust sublimation, 
$2000\K\lesssim T \lesssim 3000\K$, 
we approximate
\begin{equation}
\langle Q\rangle_T \approx 
\frac{A a_{-5} (T/2300{\rm K})}{1+Aa_{-5}(T/2300{\rm K})}
\label{eq:Q_T}
\end{equation}
with $A\approx0.03$ and 0.3 for astronomical silicate and graphite,
respectively. We adopt equation (\ref{eq:Q_T}) with
$A=0.1$ as representative of refractory grain material.

Radiation and sublimation then contribute equally to the
cooling at a temperature $T_{\rm r=s}$ determined by
\begin{equation}
T_{\rm r=s} = 2820\K
	\left\{
	1-0.040\ln\left[\left(\frac{T_{\rm r=s}}{2800\K}\right)^4
	\frac{a_{-5}(T_{\rm r=s}/2300\K)}
	{1+0.1a_{-5}(T_{\rm r=s}/2300\K)}\right]
	\right\}^{-1} ~~~.
\label{eq:T_r=s}
\end{equation}

For $T>T_{\rm r=s}$, 
so that sublimation cooling dominates over radiative cooling, 
the grain temperature is 
\begin{equation}
T_{\rm sub}\approx 3030{\rm K}\left[1+0.043\ln(
\Qabs
e^{-\tau}
L_{49}
r_{19}^{-2})\right]\quad,
\label{eq:T_sub}
\end{equation}
where $r\equiv 10^{19}r_{19}\cm$.

For $T<T_{\rm r=s}$ radiative cooling dominates and the grain
temperature is 
\begin{equation}
T_{\rm rad}=\left(e^{-\tau}
{L_{1-7.5}
\over 16\pi\sigma r^2}
{\Qabs
\over\langle Q\rangle_T}\right)^{1/4}
=
2160{\rm K}
\left(e^{-\tau}
{L_{49}\Qabs
\over\langle Q\rangle_T/0.1}\right)^{1/4}
\left({r_{19}\over 4}\right)^{-1/2}\quad.
\label{eq:T_rad}
\end{equation}
where $e^{-\tau}$ is the instantaneous
attenuation of the 
1--7.5~eV
flash by intervening absorption.

If attenuation by intervening material can be neglected,
grains are completely sublimed out to a destruction radius 
$R_{\rm d}=R_{\rm c}$
where $R_{\rm c}$ is the radius where the unattenuated flash can heat
grains to the critical temperature $T_c$.
For our nominal parameters, Eq. (\ref{eq:T_c}) and (\ref{eq:T_r=s}) give
$T_c < T_{r=s}$, so that
radiative cooling dominates at $r=R_{\rm c}$,
and the critical distance $R_{\rm c}$ is given by
\begin{equation}
R_{\rm c} =
\left(
\frac{L_{1-7.5}}{16\pi\sigma T_c^4}
\frac{\Qabs}{\langle Q\rangle_{T_c}}
\right)^{1/2}
\approx 3.7\times10^{19}
\left(\frac
{\Qabs L_{49}
(1+0.1a_{-5})}
{a_{-5}}\right)^{1/2}
\cm
\quad.
\label{R_c}
\end{equation}
In the optically-thin limit, then, the optical-UV flash from the GRB
will destroy dust {\it in the beam} out to a substantial distance.

\subsection{Effects of Dust Optical Depth}

While the 
1--7.5~eV photons from a
GRB may be capable of destroying dust out to
$R_{\rm c}\approx10$~pc
in the optically-thin limit, in dusty regions (such as molecular clouds)
attenuation of the radiation by dust
grains before they are destroyed will limit the grain destruction to
a smaller radius. 
The dust optical depth is a function
$\tau(r,t)$ of both space and time, as the GRB flash ``burns'' its way
through the cloud.

The attenuation of $\sim$1--7.5~eV photons 
is dominated by dust.
To estimate the effect of high optical depth, we will assume
that only a single dust type is present.
Let $R_{\rm d}$ be the dust destruction radius: 
all grains at $r < R_{\rm d}$ are
destroyed by the heating effects of the optical-UV flash.

We approximate the 1--7.5~eV emission from the GRB as a rectangular
pulse.
At radii $r<R_{\rm d}$, the {\it leading} edge of the optical-UV pulse is
attenuated by the dusty medium through which it propagates, but the
{\it trailing} edge of the pulse is unattenuated since it propagates through
a dustless medium, and we are neglecting the effects of gas-phase
absorption.
Rather than solve for $\tau(r,t)$, we will simplify the problem
by assuming that the effects of extinction can be approximated
as primarily a {\it narrowing} of the optical pulse, retaining a rectangular
profile.
The problem then reduces to determination of a function $f(r)$,
the fraction of the flash energy which is absorbed by dust interior to
radius $r$, and survival of grains at radius $r$ when irradiated 
by a radiation field $L_{1-7.5}/4\pi r^2$ for a
time $(1-f)\Delta t$.

The function $f(r)$ then satisfies
\begin{equation}
{df\over dr} = 
\Qabs n_d \pi a^2 
{\min[t_{\rm surv},(1-f)\Delta t]\over \Delta t}
\end{equation}
where $t_{\rm surv}(r)$ 
is the survival time of a grain irradiated by the unattenuated
radiation field.
In the unattenuated portion of the optical pulse, grains are heated to 
temperatures given by equations (\ref{eq:T_sub},\ref{eq:T_rad}) with
$\tau=0$.
The dust destruction radius $R_{\rm d}$ is then determined by the condition
$t_{\rm surv}(T(R_{\rm d})) = [1-f(R_{\rm d})]\Delta t$.

Figure \ref{fig:R_d} shows $R_{\rm d}$ as a function of cloud density
$n_{\rm H}=n({\rm H}) + 2n({\rm H}_2)$ for a standard dust-to-gas ratio
$n_d (4\pi/3)a^3\rho/n_{\rm H}m_{\rm H} = 0.01$,
for several different values of the GRB 1--7.5~eV luminosity
($L_{49}$), duration ($\Delta t_1$), and characteristic dust
grain radius ($a_{-5}$).
The radius $R_d$ for ``typical'' GRB parameters is shown as the
heavy curve.
We see that if a GRB occurred in a dusty region, 
the optical-UV flash from the GRB can clear out
a substantial amount of 
dust which lies in the beam.

\subsection{H and H$_2$ Absorption}

Radiation with $h\nu>13.6$~eV may ionize H and H$_2$.
Neglecting dust opacity, the prompt flash will be able to photoionize
a mass 
\begin{equation}
M_{\rm ion}\approx \frac{m_{\rm H} L_{13.6-50}\Delta t}{25\eV}\approx 
4\times10^3 M_{\sun} 
L_{49}\Delta t_1
\end{equation}
or out to a radius
\begin{equation}
R_{\rm ion} \approx 1.2\times10^{19}\cm\left(
\frac{L_{49}
\Delta t_{1}}
{n_3}\right)^{1/3}
\end{equation}
where $n_3\equiv n_{\rm H}/10^3\cm^{-3}$, and we approximate
the flash by a rectangular pulse of duration
$\Delta t = 10
\Delta t_{1}
{\rm\, s}$.
We have taken the typical 13.6--50~eV luminosity to be
$\sim2\times 10^{49}L_{49}\erg\s^{-1}$.

The dust destruction radius, $R_d$, is compared with $\Rion$ in 
Figure \ref{fig:R_d}. At densities for which $R_d>\Rion$, 
typically $n_{\rm H}\gtrsim 10^2\cm^{-3}$ (characteristic of
star-forming regions), our neglect of absorption of $h\nu > 13.6\eV$
photons by dust in estimating $\Rion$ is justified,
because the dust will be destroyed by absorption of
$h\nu < 7.5\eV$ photons prior to arrival of most of the ionizing
photons at a given point in the cloud. At these densities,  
$h\nu>13.6\eV$ photons are fully
absorbed by the gas in a region smaller than the dust destruction zone,
thus justifying our neglect of ionizing radiation when estimating $R_d$.
Since the number of 7.5--13.6~eV photons is comparable to the number
of $h\nu > 13.6\eV$ 
photons, this also shows that, for densities $n_{\rm H}\gtrsim10^2\cm^{-3}$,
the 7.5--13.6~eV photons will be mainly absorbed by H$_2$, and can
be neglected for purposes of dust destruction.

At lower densities, where $R_d<\Rion$, some fraction of $h\nu>7.5$~eV radiation
would also contribute to dust sublimation. At these densities 
the grain temperature near $R_d$ is determined
by radiative cooling and therefore
$R_d\propto L^{1/2}$, where $L$ is the UV luminosity available for dust
heating. Since $L_{7.5-50}\simeq2L_{1-7.5}$, the contribution of 
$h\nu>7.5$~eV radiation may increase $R_d$ by up to $\simeq50\%$.

\section{Electrostatic Disruption?}

Because of the large fluence of energetic photons, 
dust destruction could, in principle, also result from extreme ionization
of the dust grain.
High degrees of ionization could result in fission of the grain,
or emission of individual ions from the grain surface by the process 
known as ``ion field emission''.

\subsection{Coulomb Explosions?}

For an approximately spherical grain of radius $a$, charged to a potential
$U$, the tensile stress averaged over a cross section $\pi a^2$ is
$S=(U/a)^2/4\pi$.
If the maximum tensile stress which the grain material can support is
$S_{\rm max}$, then the potential gradient and grain charge $Z$
are limited by
\begin{equation}
\left(\frac{U}{a}\right) < 1.06\times 10^8 
\left(\frac{S_{\rm max}}{10^{10}{\rm dyne}\cm^{-2}}\right)^{1/2}{\rm V}\cm^{-1}
\end{equation}
\begin{equation}
Z < 7.4\times10^4 \left(\frac{S_{\rm max}}{10^{10}
{\rm dyne}\cm^{-2}}\right)^{1/2}
a_{-5}^2
\end{equation}
We note that if a grain with $S=S_{\rm max}$ fissioned into two halves,
each of the fragments would have $S \approx 2^{-1/3}S_{\rm max}$ and there
would therefore be no further fragmentation unless additional ionization
took place.

A grain contains $\sim 3\times10^9 a_{-5}^3$ electrons.
If the mean atomic number
is $\sim10$ and the photoionization cross section per electron is
$\bar{\sigma}\approx 10^{-24}\cm^2$, then substantial grain
destruction by this process would require a fluence
$F(h\nu > 10{\rm keV})\gtrsim 2.5\times10^{19}a_{-5}^{-1}\cm^{-2}$. 
Grain fission would therefore occur within a
radius
\begin{equation}
R_{\rm Fis} \approx 30 {\rm pc} E_{53}^{1/2} a_{-5}^{1/2}
\end{equation}
where we have taken the fluence 
$F(h\nu > 10{\rm keV})\approx (E/20{\rm keV})/4\pi R^2$.
Note that this process changes the grain size distribution and therefore
affects the optical extinction curve, but grain fission alone would
not appreciably reduce the ultraviolet extinction, and might even increase it.

\subsection{Ion field emission}

Ideal materials have tensile strengths 
$S_{\rm max}\approx 10^{11}{\rm dyne}\cm^{-2}$, so that a Coulomb explosion
would not take place until the surface electric field reaches
$U/a\approx 3\times10^8{\rm V}\cm^{-1}$.
However, in the laboratory electric fields exceeding
$\sim 3\times 10^8{\rm V}\cm^{-1}$ are observed to
result in ``ion field emission'', where individual ions are emitted
from the sample
(Muller \& Tsong 1969).
As a result, if the grain tensile strength $S_{\rm max}\gtrsim10^{11}{\rm dyne}
\cm^{-2}$, intense irradiation by $h\nu\gtrsim10{\rm keV}$ photons
will first cause the grain to charge up to
$Z_{\rm max}\approx 2.1\times10^5 (U/a)_{\rm max}$, and each subsequent
ionization will result in emission of an ion (assuming that electron
capture is negligible during the $\sim 10$~s of the
gamma ray burst).

If the mean atomic number is $\sim10$ and the
photoionization cross section per electron is
$\bar{\sigma}\approx 10^{-24}\cm^2$, then substantial grain
destruction by this process would require a fluence
$F(h\nu > 10{\rm keV})\gtrsim 10^{23}\cm^{-2}$. 
Grain destruction by ion field emission would therefore occur only within a
radius
\begin{equation}
R_{\rm IFE} \approx 0.5 {\rm pc} E_{53}^{1/2}
\end{equation}
Ion field emission is evidently much less important than
sublimation.

\section{Dust Infra-Red Emission}

For clouds of mass $M<10^7M_\odot$ extending to $r\gtrsim8$~pc,
most of the optical-UV energy absorbed by sublimated dust
is re-radiated in the infra-red, typically around $\lambda=1\mu$m. 
This is due to the fact that 
at distances larger than $R_{\rm r=s}\approx8L_{49}^{1/2}$~pc
radiative cooling of dust grains dominates
over sublimation cooling [see Eqs. (\ref{eq:T_rad},\ref{eq:T_r=s})], and
to the fact that for $M<10^7M_\odot$ grains are heated to the
temperature $T_c\approx2300$~K required for complete sublimation 
out to a distance $R_{\rm d}\approx10$~pc [see Eq. (\ref{eq:T_c}), Fig. 1]
(For $M>10^7M_\odot$, optical photons are completely 
absorbed at distances $\ll10$~pc, where sublimation cooling dominates and
only a small fraction of the absorbed energy is re-radiated). 

At radii where radiation cooling dominates, the grain temperature 
drops approximately as $T\propto r^{-1/2}$.
The strong dependence of grain survival time $t_{\rm surv}$
on temperature, Eq. (\ref{eq:t_sur}), then implies a sharp transition, 
i.e. over a distance $\Delta r\ll R_{\rm d}$, between the region at 
$r<R_{\rm d}$ where $t_{\rm surv}$ is much smaller than the flash
duration $\Delta t$, to the region at $r>R_{\rm d}$, where 
$t_{\rm surv}\gg\Delta t$. Since the energy radiated by grains is
proportional to $\max(t_{\rm surv},\Delta t)$, infra-red emission
of sublimated dust is dominated by emission from grains just 
outside $R_{\rm d}$. If the optical depth for UV photons due to dust at 
$r>R_{\rm d}\approx10{\rm pc}$
is $\tau_{\rm UV}\gtrsim1$, then most of the flash energy would be absorbed by
dust and re-radiated in the infra-red. Most of the infra-red radiation
would escape the cloud, and may therefore be detected, if the
infra-red optical depth is not high, $\tau_{\rm IR}\lesssim1$. For
$Q_{\rm abs,\nu}\propto\nu^1$, the requirements $\tau_{\rm UV}\gtrsim1$ and
$\tau_{\rm IR}\lesssim1$ may be written as $0.2\lesssim\tau_V\lesssim2$.

In order to estimate the dust infra-red luminosity in the case
where $\tau_{\rm IR}\lesssim1$ and $\tau_{\rm UV}\gtrsim1$, 
let us first assume that optical-UV flash emission
is beamed into a small solid angle around the line of sight,
$\Delta\Omega=\pi\theta^2\ll4\pi$. The observed duration of the
infra-red emission is then $\Delta t_{\rm IR}\approx R\theta^2/2c$, 
where $R\sim10$~pc is the radius out to
which grains are heated to $\approx2300$~K. This may be written
in the form 
$\Delta t_{\rm IR}=2(R/c)(\Delta\Omega/4\pi)\approx20
(\Delta\Omega/0.01)$~day, which is valid
for $\Delta\Omega=4\pi$ as well as for $\Delta\Omega\ll4\pi$. The infra-red
luminosity is given by the ratio of the flash energy absorbed by dust,
$E_{1-7.5}\approx10^{50}(\Delta\Omega/4\pi)$~erg, and the observed duration
$\Delta t_{\rm IR}$, $L_{\rm IR}\approx E_{1-7.5}/\Delta t_{\rm IR}\approx
10^{41}{\rm erg/s}$.

\section{Implications}

\begin{figure}
\plotBTD{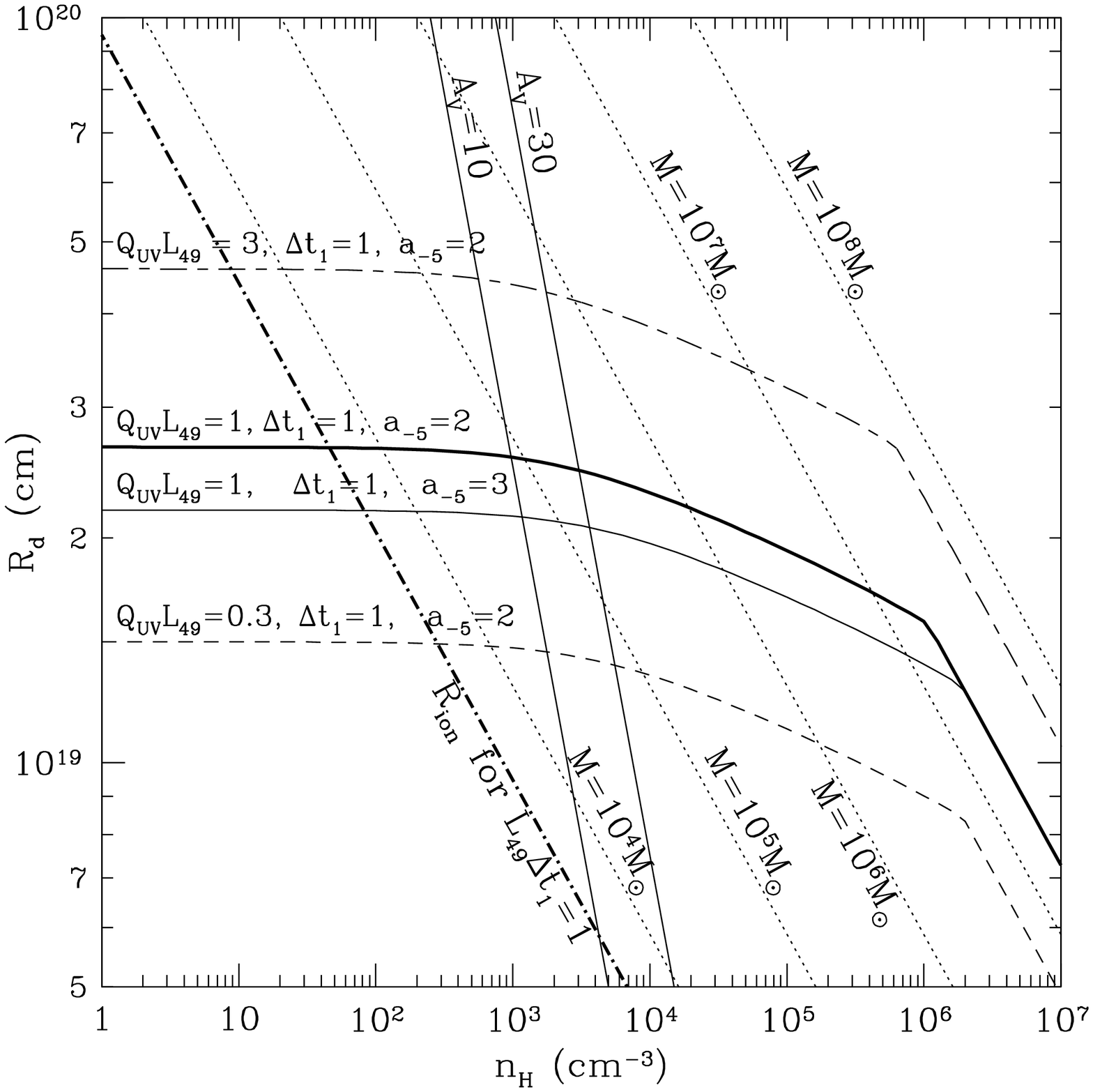}{6.0in}
\figcaption{Radius $R_{\rm d}$ out to which grains are destroyed by thermal
	sublimation, as a function of cloud density $n_{\rm H}$, for
	different values of $L_{49}$, $\Delta t_1$, and $a_{-5}$.
	The heavy curve is for ``typical'' GRB parameters.
	The lines $A_V=10$ and $A_V=30$ show the radius of a cloud
	having $A_V=10$ and 30 from center to edge.
	Also shown (broken line) is the radius $R_{\rm ion}$ out to which
	the gas is photoionized by a flash with $L_{49}\Delta t_1=1$.
	Dotted lines indicate radii with enclosed gas mass from 
	$10^4 M_{\sun}$ to $10^8 M_{\sun}$.
	\label{fig:R_d}}
\end{figure}

The luminosity of the prompt optical-UV emission accompanying GRB
$\gamma$-ray emission is given by Eqs. (\ref{L_nu}) 
and (\ref{Delta_t}).
For typical GRB parameters 
we expect an optical-UV flash with 1--7.5~eV luminosity
$L_{1-7.5}\approx1\times10^{49}{\rm erg/s}$,
assuming isotropic emission.
Such a UV flash can destroy dust by sublimation
out to an appreciable distance, 
$R_{\rm d}\approx10$~pc 
(see Figure \ref{fig:R_d}), and may
clear the dust out of $\sim 10^7(\Delta\Omega/4\pi) M_\odot$ of molecular 
cloud material, where $\Delta\Omega$ is the solid angle into which 
the optical-UV emission is beamed, and where dust is sublimed.
If GRB sources indeed lie in dusty regions, then the extinction
would decrease with time during prompt optical-UV emission, over tens of 
seconds. Detection of such time 
dependent extinction would provide strong constraints on the 
GRB environment. The destruction of dust implies that existing, or
future, observations of not-heavily-reddened fireballs are not
inconsistent with GRBs being associated with star formation.

We have shown in \S5 that if the optical depth due to dust 
beyond $\approx8$~pc
is of order unity, 
most of the UV flash energy is absorbed and re-radiated
in the infra-red, typically at 
$\lambda\approx1\mu$m. The resulting infra-red luminosity,
$L_{\rm IR}\approx10^{41}{\rm erg/s}$, extends over
an apparent duration of $\approx20(1+z)(\Delta\Omega/0.01)$~day. 
For GRBs at $z=1$, therefore, K-band photometry may reveal 
thermal emission from dust grains. 

In fact, such emission may already have been observed in 
GRB970228 (\cite{Fruchter99}) and GRB980326 (\cite{Bloom99}). 
In both cases, a deviation from a power law decline of optical flux,
which at early time is consistent with synchrotron emission from shock
accelerated electrons,
is observed at $\sim30$~d delay. As the flux drops below $\sim1\mu$Jy,
a new infra-red emission component is revealed, with a flux 
$f_\nu\approx0.5\mu$Jy between $\lambda=2\mu$m and $\lambda=1\mu$m for
GRB970228 and $f_\nu\approx0.7\mu$Jy at $\lambda=0.9\mu$m for
GRB980326. In both cases, the spectrum is modified at this time to
$f_\nu\propto\lambda^3$ at $0.5\mu{\rm m}\lesssim\lambda\lesssim0.9\mu$m.
The infra-red flux is of the same order of magnitude estimated for dust
grain emission, and the spectrum is consistent with dust emission
peaking at $\approx1\mu$m (at the source redshift), provided GRB980326
is at redshift $z\sim0.4$. We note that for the typical parameters 
adopted in this paper, dust emission is expected to peak at somewhat
longer wavelength, $\approx1.5\mu$m. However, since the grain properties
are not well known, dust emission can not be ruled out as an alternative
to the 
proposal that the ``excess'' emission is due to a supernova 
(\cite{Bloom99,Reichart99,Galama99})
We note also that 
the non-detection of optical emission from GRB980326 at $t\sim200$~d,
implies, under the dust emission hypothesis, beaming of the
optical-UV flash to $0.01\lesssim\Delta\Omega\lesssim0.1$, consistent
with the interpretation that the rapid, $t^{-2}$, flux decline is
due to the fireball being a jet of small opening angle (\cite{Rhoads99}).

We have shown in \S2 that strong optical-UV emission requires,
like GRB $\gamma$-ray production, large initial expansion
Lorentz factor, $\Gamma_i>100$, which also implies that
the plasma emitting the optical-UV flash must be expanding with 
$\Gamma>100$. Thus, if 
the fireball is a jet of finite opening angle,
$1\gg\theta_j>1/\Gamma\sim0.01$, 
then both $\gamma$-ray and optical-UV emission will be confined to a small 
solid angle $\Delta\Omega=\pi\theta_j^2$.
In this case, dust would be evaporated only within a narrow cone 
around the line of sight.
A jet-like fireball expands as a conical section of a spherical
fireball, with $\Gamma\propto t^{-3/8}$, as
long as $\Gamma>1/\theta_j$. After deceleration to $\Gamma<1/\theta_j$,
the jet expands sideways, its opening angle increasing to $\simeq1/\Gamma$
and 
$\Gamma\propto t^{-1/2}$ (\cite{Rhoads99}). At this stage 
radiation reaching us must travel a distance 
$l\approx r\sin(\Gamma^{-1}-\theta_j)/\sin(\theta_j)$
through gas which was not exposed to the initial flash, and which will still
contain dust. For $1\ll\Gamma\ll1/\theta_j$, and using
$t\approx r/2\Gamma^2c$, we have
$l\approx r/\Gamma\theta_j\approx2\Gamma ct/\theta_j
\approx0.1{\rm pc}(t/1{\rm\, week})^{1/2}/\theta_j^{2/3}$.
Thus, a significant increase in extinction 
over time
would be observed if $\theta_j\ll0.1$.

Confinement of $\gamma$-ray and strong optical-UV emission to a small
solid angle is not limited to the case of a jet-like fireball. 
It may also arise if $\Gamma_i$, the initial expansion Lorentz 
factor, is anisotropic. Consider a fireball carrying similar energy per
unit solid angle in all directions, with $\Gamma_i$ a decreasing function
of angle with respect to the line of sight, such that
$\Gamma_i\ll300$ for $\theta>\theta_\gamma$. In this case, 
optical-UV (and $\gamma$-ray) emission would be suppressed at angles
$\theta>\theta_\gamma$. The isotropic fireball energy per unit solid
angle implies that, after a transition phase, the fireball would approach
spherical expansion, with $\Gamma\propto t^{-3/8}$. 
At this stage, most of the radiation detected at time $t$ by a distant
observer is produced by fireball plasma within a narrow ring of radius 
$\simeq r/\Gamma$ around the line of sight, where the fireball radius $r$ and 
expansion Lorentz factor $\Gamma(r)$ are related to $t$ through 
$t=r/2\Gamma^2c$ (\cite{Wring}). Thus, here too
a significant increase in reddening is expected once $\Gamma$
drops below $1/\theta_\gamma$. For $1\ll\Gamma\ll1/\theta_\gamma$
most of the radiation reaching us must pass through a path length
$l\approx r/\Gamma\theta_\gamma\approx2\Gamma ct/\theta_\gamma
\approx0.05{\rm pc}(t/1{\rm\, week})^{5/8}/\theta_\gamma$ of
gas which was not exposed to the initial flash.

Finally, 
it should be noted that although optical flashes are expected
for typical GRBs [with $E_{53}\approx 1$ and $n_0\approx 1$,
see Eq. (\ref{eq:f_V})], prompt optical emission has been observed for only
1 GRB to date and therefore may not accompany all GRBs.
Indeed, fireballs in low density environments with $n\ll 1\cm^{-3}$ would
not be expected to produce strong prompt emission [see Eq. (\ref{eq:f_V})].
In addition, if the fireball initial Lorentz factor $\Gamma_i\gg300$, 
reverse shock emission may be shifted to photon energies above 
7.5~eV [see Eq. (\ref{nu_m})], where most photons are absorbed by 
H and H$_2$. 

\paragraph*{Acknowledgments.} 
This work was supported in part by NSF grant AST-9619429.
EW thanks the Institute for Advanced Study, Princeton for its hospitality
during the period when this study was initiated.

\newpage
\end{document}